\documentclass[11pt,prd,superscriptaddress,nofootinbib]{revtex4}

\begin{document}

\title{\Large Anomaly and long-range forces}

\author{\bf V.P. Kirilin}
\affiliation{ITEP, B. Cheremushkinskaya 25, Moscow, 117218, Russia}
\affiliation{Department of Physics, Princeton University, Princeton, NJ 08544, USA}

\author{\bf A.V. Sadofyev}
\affiliation{ITEP, B. Cheremushkinskaya 25, Moscow, 117218, Russia}
\affiliation{Center for Theoretical Physics, Massachusetts Institute of Technology, Cambridge, MA,  02139}

\author{\bf V.I. Zakharov}
\affiliation{ITEP, B. Cheremushkinskaya 25, Moscow, 117218, Russia}
\affiliation{Max-Planck Institut f\"{u}r Physik, M\"{u}nchen, 80805, Germany}
\affiliation{Moscow Inst Phys \& Technol, Dolgoprudny, Moscow Region, 141700, Russia}

\begin{abstract}
We consider infrared dependences of chiral effects, like chiral magnetic effect,
in chiral media. 
The main observation is that there exist competing infrared-sensitive
parameters, sometimes not apparent. The value of the chiral effects depends in 
fact on the actual  hierarchy of the parameters. Some examples have been already
given in the literature. We argue that 
 magnetostatics of chiral media with a non-vanishing chiral
chemical potential $\mu_5\neq 0$ is also infrared sensitive.   
In particular, the system turns to be unstable if the volume is large enough.
The instability is with respect to the decay of the system into domains of
non-vanishing  magnetic field with non-trivial helicity. 
\end{abstract}

\keywords{Chiral Anomaly; Chiral Magnetic Effect; Infrared Effects.}

\maketitle

\section{Introduction}
Relation of the chiral anomaly to long-range interactions is actually an old topic,
with rich literature existing.  The point is that the chiral anomaly can be derived 
both in terms of ultraviolet and infrared sensitive regulators.
Namely,  divergence of 
the axial current $j_{\mu}^5$
of massless fermions of charge $e$ is given by a polynomial
in the photonic field $A_{\mu}$
\cite{adler}:
\begin{equation}\label{divergence}
\partial_{\mu}j_{\mu}^5~=~{e^2\over 8\pi^2}F_{\mu\nu}\tilde{F}_{\mu\nu}~,
\end{equation}
where $F_{\mu\nu}~=~\partial_{\mu}A_{\nu}-\partial_{\nu}A_{\mu},~
\tilde{F}_{\mu\nu}~=~1/2\epsilon_{\mu\nu\alpha\beta}F_{\alpha\beta}$.
Being a polynomial, the divergence of the current is ultraviolet sensitive 
\cite{adler}. 

On the other hand, if one turns to the matrix element of the axial currrent
itself, $\langle\gamma\gamma|j_{\mu}^5|0\rangle$ it exhibits a pole \cite{dolgov}:
\begin{equation}\label{pole}
\langle\gamma\gamma|j_{\mu}^5|0\rangle~=~{i q_{\mu}\over q^2}{e^2\over 2\pi^2}
\epsilon_{\rho\nu\alpha\beta}
k^{(1)}_{\rho}k^{(2)}_{\nu}e^{(1)}_{\alpha}e^{(2)}_{\beta}~,
\end{equation}
where $q_{\mu}$ is the 4-momentum brought in by the axial current,
$k^{(1)}_{\alpha},~k^{(2)}_{\beta}$ and $e^{(1)}_{\mu},e^{(2)}_{\nu}$
are the 4-momenta and wave functions of the photons, respectively. 
Note that Eq. (\ref{pole}) is only valid for photons on mass shell,
$(k^{(1)})^2=(k^{(2)})^2~=~0$.
The emergence of the pole in $q^2$ in Eq. (\ref{pole}) is a pure perturbative
 phenomenon and a reflection of the vanishing
fermionic mass.

However, the pole is to be exhibited also in the massless limit of strongly interacting
quarks when the perturbation theory does not apply.
 In a confining theory this requirement results in
the 't Hooft matching condition \cite{thooft}:
\begin{equation}
{e^2\over 8\pi^2}N_c(Q_u^2-Q_d^2)~=~f_{\pi}\cdot f_{\pi\to \gamma\gamma}~,
\end{equation}
where $N_c$ is the number of colors, $Q_u=2/3, Q_d=-1/3$ are 
the quark charges
and $f_{\pi},f_{\pi\to \gamma\gamma}$
are the constants related to the pion decays into a lepton pair and two photons,
respectively.
 
Thus, within this picture the divergence of the axial current is driven by 
the ultraviolet
physics alone 
while the matrix elements of the current are infrared sensitive. However,
this  is correct only in the limit of  exact chiral symmetry and in this sense
is an oversimplification. Indeed, if one introduces small but finite quark masses
the matrix element of the divergence of the current becomes also infrared
sensitive:
\begin{eqnarray}\label{me}
\langle\gamma\gamma|\partial_{\mu}j^5_{\mu}|0\rangle~\approx~
f_{\pi}f_{\pi\to \gamma\gamma}\cdot 
\Big(1~-~{m_{\pi}^2\over (q^2+m_{\pi}^2)}\Big)F_{\mu\nu}\tilde{F}_{\mu\nu}~,
\end{eqnarray}
where $m_{\pi}^2$ is the pion mass squared.
We see that the result now depends strongly on the ratio of the
two infrared sensitive parameters involved, $q^2$ and $m_{\pi}^2$.
The exclusively ultraviolet-sensitive origin of the
non-trivial divergence of the current, see Eq. (\ref{divergence})
appears to be property of a particular hierarchy of the infrared-sensitive
parameters, $q^2~\gg~ m_{\pi}^2$. In the opposite limit of $m_{\pi}^2~\gg~q^2$
the matrix element (\ref{me})  is vanishing.

The actual focus of our attention is the infrared sensitivity
of chiral effects in chiral media, that is media whose constituents are massless
fermions. The chiral media attracted a lot of attention recently,
for a review and further references see, e.g., \cite{review}.
 One of main reasons is that one expects that in such media
the chiral anomaly (\ref{divergence}) which is a pure loop, or quantum
effect has macroscopic manifestations. The best known example
of such manifestations  is
the chiral magnetic effect \cite{kharzeev}:
\begin{equation}\label{cme}
j_{\mu}^{el}~=~\sigma_M\mu_5 B_{\mu}~,
\end{equation}
where $j_{\mu}^{el}$ are components of the electromagnetic current,
$\mu_5$ is the chiral chemical potential,
$\sigma_M$ can be called magnetic conductivity,
while $B_{\mu}$ is defined in terms of external electromagnetic field
$F_{\alpha\beta}$ and four-velocity $u_{\nu}$ of an element of
the chiral liquid, 
$B_{\mu}\equiv~(1/2)\epsilon_{\mu\nu\alpha\beta}u_{\nu}F_{\alpha\beta}$. 
Moreover, $\sigma_M$
is fixed \cite{surowka} in the hydrodynamic approximation by
the chiral anomaly
(\ref{divergence}) and equals to:
\begin{equation}\label{mu1}
\sigma_M~=~\frac{e^2}{2\pi^2}~,
\end{equation}
where $e$ is the electric charge of the constituents.

We will discuss also the parity-reflected cousin 
of Eq (\ref{cme}): 
\begin{equation}\label{separation1}
j_{\mu}^5~=~\sigma_5 \mu \cdot B_{\mu} ~,
\end{equation}
where $\mu$ is the chemical potential,
conjugated to the electric charge.
According to \cite{metlitski} the coefficient $\sigma_5$ is again
uniquely determined in terms of the anomaly:
\begin{equation}\label{separation2}
\sigma_5~=~\frac{e}{2\pi^2}.
\end{equation}
Another effect which we have in mind is the so called chiral vortical effect:
\begin{equation}
j^5_{\mu}~=~\frac{1}{2}\sigma_{\omega}\epsilon_{\mu\nu\alpha\beta}u_{\nu}
\partial_{\alpha}u_{\beta}~,
\end{equation}
where $\sigma_{\omega}$, to the lowest order in the chemical potentials $\mu, \mu_5$ 
can be expressed, again, in terms of the anomaly
\cite{surowka,jensen}:
\begin{equation}
\sigma_{\omega}~=~\frac{\mu^2+\mu_5^2}{2\pi^2}~+~O(\mu_{(5)}^3).
\end{equation} 
As for the corrections of order $O(\mu_{(5)}^3)$ they also arise 
in the hydrodynamic approximation, generally speaking, see, e.g., 
\cite{surowka}. However, these corrections are sensitive to details of
infrared regularization, see, e.g., \cite{shevchenko}, and can be removed by 
proper choice of the coordinate frame.  

All the chiral effects we mentioned refer to equilibrium, as first emphasized in Ref.
 \cite{cheianov}.
As a result, the linear-response  relation for the transport coefficient
$\sigma_M$ looks so as if we were considering a static effect:
\begin{eqnarray}
\label{sigmaCME}
\sigma_M ~= ~\lim_{k_n\to 0} \epsilon_{ijn} \frac{i}{2 k_n} \langle j_i,j_j\rangle,
\end{eqnarray}
where $\langle j_i,,j_j\rangle\equiv\Pi_{ij}$
is the current-current correlator in the momentum space.
Similarly, for the chiral separation effect one gets:
\begin{equation}\label{separation}
\sigma_5 ~=~ \lim_{k_n\to 0} \epsilon_{ijn} \frac{i}{2 k_n} \langle j^{(5)}_i,j_j\rangle~.
\end{equation}
 For derivation of Eqs (\ref{sigmaCME}), (\ref{separation})
see, e.g., \cite{kharzeev,jensen,landsteiner} and references therein.
Note that in case of  the standard Kubo formulae
the limiting procedure is different.
Namely, in that case the spatial momentum is vanishing identically,
 ${k}_i\equiv  0$ while the frequency tends to zero, $\omega \rightarrow 0$.   
 
Eq. (\ref{sigmaCME}) is a convenient starting point
  to explain, what kind of problems we will address
here.
Eq. (\ref{sigmaCME}) relates the chiral conductivity to a polynomial in the current-current
correlator, $\Pi_{ij}\sim~\epsilon_{ijn}k_n$. The polynomial, in turn, is determined by
ultraviolet-sensitive regulators.  One can argue, therefore, that it is crucial to regularize
the product of the two electromagnetic currents, entering $\Pi_{ij}$ at coinciding points. 
A detailed derivation of $\sigma_M$ along these lines can be found in  Ref.  \cite{landsteiner}
and does reproduce the standard result (\ref{mu1}).
Note that the ultraviolet behavior of the correlator of the currents
is determined by the subtraction constant and fixed uniquely by the theory.

We feel, however, that ascribing  the chiral magnetic effect entirely
to the ultraviolet physics
 is an oversimplification. Namely, in analogy with the discussion above
 we expect that there exist alternative derivations of the chiral effects
which are sensitive to infrared physics. Moreover, we expect, that the final
result depends on hierarchy of infrared-sensitive parameters, see for a discussion
above. To uncover such a hierarchy it is useful to introduce again 
(compare  Eq. (\ref{me}))
a finite
fermion mass, violating the chiral symmetry. Then one can argue \cite{zakharov2}
that the standard value of the chiral conductivity (\ref{mu1}) corresponds
to the following hierarchy of the infrared-sensitive parameters:
\begin{equation}\label{inequality}
e|k|~\gg~e\omega~ \gg ~m_f~.
\end{equation}
Further examples of infrared dependences of the chiral magnetic effect  
 can be found in Ref. \cite{chinese} .

\section{From a non-local function to a polynomial}

At first sight, the non-local expression (\ref{pole}) for the matrix element of an anomalous current 
is very different
from polynomials which, according to (\ref{sigmaCME}), (\ref{separation})
 determine the chiral effects.
As a preliminary remark, we argue in this section that in some, physically motivated limits the
non-local functions do reduce to polynomials.

As first example, consider quantum chromodynamics at
finite temperature in the  Euclidean formulation of the theory.
Moreover, consider the limit of exact chiral symmetry. The chiral anomaly
at finite temperature has been considered, of course, in many papers,
see for example \cite{pisarski}. Here we consider only the resolution of an apparent contradiction
between existence of a massless pion and absence of long range-forces
at finite temperature.

In more detail, 
as far as the temperature is small compared to the temperature of the
deconfining phase transition $T_c$ the chiral symmetry remains spontaneously broken and
there is a massless pion. Thus, we expect the pole (\ref{pole}) 
to be present in the matrix elements
of the axial current. On the other hand, the 4d theory reduces now to a sequence of
3d theories with finite fermion masses:
\begin{equation}
m_f^{(3d)}~=~2\pi T\big(n+1/2\big)~,
\end{equation}
This implies that perturbatively the triangle graph does not correspond any longer
to an infinite-range interaction, in an apparent contradiction with existence of the
pole (\ref{pole}). This is the paradox we would like to address.

Let us consider the kinematics in more detail. The extension of space
in the time direction is now
limited to $$\tau~\le~1/T~,$$
where $\tau$ is the Euclidean time coordinate. Thus, one can probe propagation to large
distances only in spatial directions. 
 Moreover, the
pion is massless only in case of the 3d theory corresponding
to the Matsubara frequency $\omega_M=0$ and we concentrate on this case.
We can choose  $q_i~\sim~(0,0,q_3)$ and rewrite the 
non-trivial component of the matrix element (\ref{pole}) as:
\begin{eqnarray}\label{common}
\langle j_3^{(5)}\rangle_{external ~fields}~=~ {q_3q_3\over q_3^2}{e^2\over 2\pi^2}\epsilon_{3ij}A_0 ik_iA_j
=~{e\mu\over 2\pi^2}B_3~
\end{eqnarray}
where we replaced the external  potential $e A_0$ by the chemical potential $\mu$
and $B_3$ is the external magnetic field. Clearly,  
we succeeded to rewrite Eq. (\ref{me})  in the form which 
coincides with the equation (\ref{separation1})
for  the chiral separation effect.

The central point of this simple exercise is that the pole exhibited in Eq (\ref{me}) disappears
in case of the specific kinematics relevant to the chiral effects. The reason is that
we do not observe Lorentz covariance any longer since the rest frame of the liquid is singled
out on physical grounds. 
The cancellation of the pole we observed is universal for any matrix 
element associated with the pion exchange.
Indeed, it is a general property of interactions of Goldstone particles that at small momenta
all the vertices are proportional to the momentum of the massless particle. And if the momentum
of the pion has only a single non-vanishing component the cancellation is obvious.

Note also that we can readily read off  Eq (\ref{separation}) from Eq. (\ref{common}).
In this sense, the both approaches are equivalent to each other
 in case of exact chiral symmetry.
However, if we start from the non-local equation (\ref{me}) we are better equipped to study
dependence on extra infrared-sensitive parameters. In particular, turning on a finite pion mass
turns off the chiral separation effect at distances $d~\gg ~m_{\pi}^{-1}$:
\begin{equation}
\lim_{q_3/m_{\pi}\to 0}{\langle j_3^{(5)}\rangle}~ =~0~,
\end{equation}
in analogy with  Eq. (\ref{me}).
 
 Another example of this type is provided by evaluation of matrix element  
of the  axial charge over a
photon state:
\begin{equation}\label{mee}
Q^{axial}_{\gamma}~~=~\langle\gamma|\hat{Q}^{axial}|\gamma\rangle, ~~\hat{Q}^{axial}~=~
\int d^3x~\bar{\Psi}\gamma_0\gamma_5\Psi~.
\end{equation}
Definition of the charge assumes that
$$\vec{q}~\equiv~0, ~~q_0~\to~0~,$$
where $q_{\mu}$ is the momentum carried in by the axial current.
The 4-momentum of the photon, $k_{\nu}$ can be arbitrary, on the other hand.
We can use now the result (\ref{pole}) for the matrix element (\ref{mee}):
\begin{equation}\label{axial}
Q^{axial}_{\gamma}~=~-i\frac{e^2}{8\pi^2}\frac{q_0}{q_0^2}F_{\mu\nu}\tilde{F}_{\mu\nu}~=~
\frac{e^2}{4\pi^2}\epsilon_{0ijk}A_i\partial_jA_k~,
\end{equation}
where $A_i$ is the vector potential. We see again that, 
once we do not impose Lorentz covariance, the non-locality  
vanishes in the chiral limit and we come to a polynomial.

Eq. (\ref{axial}) implies that non-vanishing axial charge is to
be ascribed to certain configurations of external magnetic fields.
More specifically, let  us introduce helicity of electromagnetic field as
\begin{equation}\label{helicity}
\mathcal{H}=\int \vec{A}\cdot \vec{B} d^3x~.
\end{equation}
Then the chiral anomaly implies, by passing from Fourier to coordinate space, that we have to ascribe axial charge 
\begin{equation}\label{charge}
Q^{axial}_{\gamma}~=~\frac{e^2}{4\pi^2}\mathcal{H}
\end{equation}
to classical magnetic field configuration.

\section{Evaluation of the chiral effects}

In the presence of  a chemical potential $\mu_5\neq 0$ the Hamiltonian of the system, 
$H_0$ is redefined as
$$H_0~\to~H_0~-~\mu_5Q^{axial}~~.$$
Moreover, in the Lagrangian language $\delta L~=~-\delta H$. Eq (\ref{axial}) 
then implies a modification
of the effective action:
\begin{equation}\label{action1}
\delta S_{eff}~=~\int dt d^3x~\mu_5\frac{e^2}{4\pi^2}\epsilon_{ijk}A_i\partial_jA_k~ .
\end{equation}
Electromagnetic current can be evaluated by varying the effective action with respect to 
the potential $A_i$. As a result we get, as is expected:
\begin{eqnarray}
{\bf j}_{el}~=\mu_5\frac{e^2}{2\pi^2}{\bf B}~,
\end{eqnarray}
where ${\bf B}$ is the external magnetic field.

The simplest way \cite{shevchenko} 
to derive the chiral vortical effect is to utilize the analogy between
gauge potential $A_{\mu}$ of field theory and local 4-velocity of an element of liquid
$u_{\mu}$: 
\begin{equation}\label{substitution}
e\cdot A_{\mu}~\to~\mu\cdot u_{\mu}~.
\end{equation}
Indeed in case of liquid the effective interaction is given by:
\begin{equation}
L_{int}^{eff}~=~\mu u_{\mu}\bar{\Psi}\gamma_{\mu}\Psi
~+~eA_{\mu}\bar{\Psi}\gamma_{\mu}\Psi~.
\end{equation}
Discussion of the validity of this analogy and references can be found in Ref. \cite{shevchenko}. It should be noted that to take into account also the $\mu_5$ contribution one has to introduce effective axial "gauge" field.

Applying substitution (\ref{substitution}) to
(\ref{charge}) , one concludes that helical macroscopic motion 
of the liquid is associated with a non-vanishing axial charge:
\begin{equation}\label{axialomega}
Q^{axial}_{\omega}~=~
\frac{\mu^2}{4\pi^2}\int d^3x~\epsilon_{ijk}u_i(x)\partial_ju_k(x)~.
\end{equation} 
An alternative derivation of the chiral vortical effect can be given in terms of the
gravitational, or mixed chiral anomaly, see, e.g.,  \cite{landsteiner} .

To summarize, both the chiral magnetic effect and chiral vortical effect have clear
connection with the chiral anomaly. Namely, the axial charge of the (massless) constituents
is not conserved because of the anomaly. However, one can introduce a conserved axial
charge by ascribing it to electromagnetic field configurations with non-vanishing helicity and
to a helical motion of chiral liquids. It is worth emphasizing that so far we assumed the chiral symmetry to be exact.  If we introduce
finite fermion mass but keep $\mu_5$ time-independent, then there is no manifestations
of the chiral anomaly. Also, the reduction of the chiral magnetic effect to the current commutator
(\ref{sigmaCME}) is lost generally speaking.

Turn now to consideration of the chiral separation effect (\ref{separation1}).
Apparently, the effect is associated with the triangle graph, generated by the
interactions 
$\mu Q^{el},~eA_{\mu}j_{\mu}^{el},~g_Aj_{\mu}^{axial}$ with the coupling $g_A=1$.
 In more detail, consider first the kinematics
similar to the one considered above, with substitution of the vertex
$\mu_5Q^{axial}$
 by $\mu Q^{el}$. Namely,
let the 3-momentum carried in by the axial current be equal to the 3-momentum carried
by the electromagnetic potential, and the vertex proportional to $\mu \cdot u_{\mu}$
be associated with 3-momentum equal to zero, $|{\bf q}|\equiv 0$ while the $q_0$
component tends to zero,
$q_0~\to~0$. Since the electric current is not anomalous,
\begin{equation}\label{zero}
Q^{el}_{\gamma_A-\gamma}~\equiv~\langle\gamma_A|\hat{Q}^{el}|\gamma\rangle~=~0~,
\end{equation}
where $\gamma_A$ is a fictitious axial photon coupled to the axial current.
From (\ref{zero})
we would conclude that
\begin{equation}
\sigma_5~=~0~.
\end{equation}
which is in apparent contradiction with evaluations of $\sigma_5$  in
\cite{metlitski}. 

One of the reasons is that our treatment of the anomaly is asymmetric
with respect to the vector and axial currents. The vector current in our case is distinguished
by its coupling to a physical massless vector particle, photon. If one concentrates on, say,
vector current associated with the baryonic quantum number, as in Ref .\cite{metlitski}, then 
the anomaly can be treated in a different way. 
For further discussion see, e.g., \cite{shevchenko}.

We could consider also another kinematics which is exactly the same as above,
with small momentum $q_0, q_0\to 0$ carried in by the axial current. Moreover we consider
now the average value of spatial components of the axial current. Then the Lorentz-covariant
completion of (\ref{axial}) would bring the result
\begin{equation}
\langle j_i^{axial}\rangle~=~\frac{\mu e}{2\pi^2}\epsilon_{ijk}\partial_jA_k~,
\end{equation}
which is equivalent to (\ref{separation1}).

For non-interacting
fermions one can also evaluate $\sigma_5$ directly, in terms of the Landau levels
\cite{metlitski}. The calculation goes through for massive fermions as well.
The only change  in $\sigma_5$ to be made is:
\begin{equation}
\mu~\to~\sqrt{\mu^2-m^2_f}~,
\end{equation} 
where $m_f$ is the fermion mass. Clearly, any relation of the matrix element $<j^{axial}_i>$
 to the chiral anomaly is lost unless $\mu$  not much larger than $m_f$.   

It follows from these remarks that, in any case,
 the status of the chiral separation effect 
(\ref{separation1}) is different
from the status of the chiral magnetic effect (\ref{cme}) \footnote{This point
was elaborated in collaboration with Sergey Vavilov.}. In the latter case we consider
axial charge and there apply various non-renormalization theorems. In the former case
we consider spatial component of the axial current and there are no reasons to expect
that any non-renormalization theorems exist. Indeed, there are no theorems on
non-renormalizability of magnetic moments of fermions. 
Also, the hierarchy of infrared-sensitive parameters for the two types of defects 
is different.  Namely, the constraint (\ref{inequality}) on the frequency $\omega$
is related to the fact that the chiral magnetic effect is associated with production
of chiral fermions. This is not true in case of the chiral separation effect. As is noted
in \cite{metlitski} for non-relativistic fermions the chiral separation  effect reduces
to evaluation of the average spin value:
$$\langle{ j^5_i}\rangle~\to~\langle{ \sigma_i}\rangle~.$$
 Thus, there is no actual flow of chirality along the magnetic field.

\section{Infrared instabilities} 

\subsection{Infrared divergences due to massless charged particles}
Chiral magnetic effect arises as a result of interplay between quantum field theory and
phenomenological, hydrodynamic approach. As far as one continues with evaluation of
quantum corrections in field theory
it is quite obvious that further infrared singularities are encountered.
Indeed, we are considering now field theory of massless charged particles which is
actually not well in defined on the mass shell and results of measurements in such a theory
would depend on the resolution, or experimental set up,
for discussion and references see, e.g., \cite{lee}.
 
On the other hand, assuming
hydrodynamics to apply, one postulates that results are not dependent, say, on the volume
of the system.
In other words, one assumes in fact
that the infrared regularization is somehow provided
by the constituents interaction, without destroying  symmetries and the
non-renormalization
theorems proven within the field theory approach, see, e.g., \cite{surowka}. 
Since there is no explicit mechanism of such an infrared cut off known, there is no guarantee
that matching of the field theory and hydrodynamics results in fact in a self-consistent
picture. There are some hints in the literature  on possible 
inconsistencies and let us mention
a few examples. 

As is expected, evaluation of the radiative correction to the chiral separation effect in
the approximation of non-interacting fermions results in an infrared unstable
expression \cite{shovkovy}:
\begin{equation}\label{shovkovy}
\delta\langle j^{axial}\rangle~=~-\frac{\alpha_{el}eB\mu}{2\pi^3}\Big(\ln \frac{2\mu}{m_f}+
\ln \frac{m_{\gamma}^2}{m_f^2}+\frac{4}{3}\Big)~,
\end{equation}
where fermion mass is taken to be $m_f\ll \mu$ and $m_{\gamma}$ is the fictitious photon mass.

The problem with the infrared divergence in the photon mass $m_{\gamma}$ 
in expression (\ref{shovkovy}) could be 
settled by considering higher orders in the magnetic field.  Then 
the fermions occupy actually the Landau levels and are off mass shell.
This would bring also $m_{\gamma}^2\neq 0$.
 It is less clear how to make sense of expression (\ref{shovkovy}) in the 
limit $m_f\to 0$. As is mentioned above, it is a highly non-trivial problem how to define
a massless charged particle on the mass shell.

Turning to the effective action (\ref{action1}), it is calculable in field theory.
Phenomenologically,
one could introduce terms which contain higher orders in derivatives.
In the hydrodynamic approximation, they are assumed to be small. 
The question is, what is the characteristic mass scale in the hydrodynamic 
expansion in derivatives. For example, 
if one considers the 2d Hall liquid, then the underlying field theory provides the scale
for the hydrodynamic approximation, which is the energy gap to the next Landau level.
If one considers a 3d liquid, then, to the best of our knowledge, there is no field-theoretic 
mechanism to provide a gap.
Nevertheless, one postulates validity of the expansions both
 in magnetic field and in the
derivatives. Under such assumptions the coefficient $\sigma_{\omega}$, for example, 
receives corrections of order $\mu^3$ \cite{surowka}:
\begin{equation}\label{mucube}
\sigma_{\omega}~=~\frac{\mu^2}{2\pi^2}\Big(1+\frac{2}{3}\frac{\mu\cdot n}{\epsilon+P}\Big)~,
\end{equation}
where $n$ is the density of particles, $\epsilon$ and $P$ are the energy density and pressure,
respectively. Note that the enthalpy, $\epsilon+P$ plays the role of a "hydrodynamic mass"
in some other cases as well.

Eq. (\ref{mucube}) refers to the temperature independent part of $\sigma_{\omega}$.
There is also contribution proportional to $T^2$, $\sigma_{\omega}^T$.
 Moreover, using the underlying field theory,
one can evaluate the first radiative correction to $\sigma_{\omega}^T$  \cite{golkar,hou}
in terms of the interaction coupling $g^2$:
\begin{equation}
\sigma_{\omega}~=~\frac{T^2}{2\pi^2}\Big(1+(const)\frac{g^2}{48\pi^2}\Big)~,
\end{equation}
where the $(const)$ depends on the fermionic representation. 
The central point is that if one considers Yang-Mills theory, the coupling $g^2$ is to be
replaced at high temperature by the running coupling $g^2(T)$. However, any logarithmic
dependence, brought by the running of the coupling, would 
not fit the thermodynamic expansion which does not reserve for any non-analyticity
in temperature.

Another reservation suggested by field theory, is that the current (\ref{cme}) would,
generally speaking, radiate. On the other hand, in classical approximation the chiral
magnetic effect is dissipation free since the current flows along the magnetic field,
and the magnetic field produces no work \cite{yee}. 

\subsection{Back-reaction of chiral medium}

In this subsection we will discuss another inconsistency between field theoretic and
hydrodynamic approaches. In hydrodynamics, one simply postulates that medium with
$\mu_5\neq 0$, placed in external magnetic field can be in equilibrium state,
and demonstrates then that there exists current (\ref{cme}). Field theory tells us that
actually such a medium is unstable. One of the explanations is that the current
(\ref{cme}) produces extra magnetic field and this effect should be taken into account.
Since this back-reaction of the medium is of next order in electromagnetic
interaction, large volumes are needed for the effect to become important. 
This instability has been discussed in Refs.
\cite{yamamoto1,khaidukov}. Moreover, similar instabilities
were discussed earlier, see, e.g., \cite{deryagin, zamaklar, volovik, shaposhnikov} in some other
sets up. We will follow the paper in Ref. \cite{khaidukov}. Moreover, 
even in the absence of external magnetic field the medium with $\mu_5\neq 0$
seems to be unstable against spontaneous generation of the magnetic field, or, probably,
domains of it.

It is convenient to start with the effective action (\ref{action1}) and 
concentrate on the static case with no dependence on time. Then, the 3d
(Euclidean) looks as:
\begin{equation}\label{action3d}
S_{eff}^{3d}~=~\int d^3x\Big(\frac{1}{2}\sigma_M\epsilon_{ijl}A_i\partial_jA_k~+eA_ij^{el}_i
-\frac{1}{4}(F_{ik})^2\Big)~.
\end{equation}
Furthermore, introduce propagator of the photon which in general
takes the form:
\begin{eqnarray}
\label{prop}
D_{ij}(\vec{k})=D_{S}(k)(\delta_{ij}-\hat{k}_{i}\hat{k}_{j})+D_{A}(k)i\epsilon_{ijl}\hat{k}^l
+ a\frac{\hat{k}_{i}\hat{k}_{j}}{k^2}~,
\end{eqnarray}
 where $\hat{k}$ is the unit vector along the 3d momentum 
 and the last term is the gauge fixing. 

Explicit summation of the bubble-type graphs associated with the 
action (\ref{action3d}) provides the following result:
\begin{equation}\label{proppole}
D_{ij}(k_i)=\frac{1}{k^2-\sigma_M^2}\Big(\delta_{ij}-\frac{k_ik_j}{k^2}\Big)
+\frac{i\epsilon_{ijl}\sigma_Mk_l}{k^2(k^2-\sigma_M^2)}~+a\frac{k_ik_j}{k^4}~.
\end{equation}
 A conspicuous feature of the expression (\ref{proppole}) is the pole in the physical region
at $k^2=\sigma^2_M$. This means instability of the perturbative vacuum with
$\langle A_i\rangle_{class}~=~0$.  
The basic assumption is the validity of the effective action (\ref{action3d}).
This action, on the hand, is commonly derived in
hydrodynamic approximation, see , e.g., \cite{jensen}.

Equation (\ref{proppole}) indicates that the true vacuum has non-vanishing magnetic field with
fixed $k^2=\sigma_M^2$. Moreover, there is also a pole in front of the structure
$\epsilon_{ijl}k^l$ which means that one expects that the magnetic field in the true
vacuum possesses non-trivial helicity (\ref{helicity}). 

These features look exotic. However, they can be readily appreciated if we look at
the action (\ref{action3d}) in somewhat different way. Namely, instead of evaluating 
propagator for the gauge field on the trivial background we can start with the
classical equation corresponding to the action (\ref{action3d}):
 \begin{equation}\label{beltrami}
{\bf curl} ~{\bf B}~=~\sigma_M {\bf B}~.
\end{equation} 
Equation (\ref{beltrami}) is nothing else but the well known Beltrami
equation, which has been studied since long in various physical frameworks,
see, in particular, \cite{beltrami}. And, indeed, the simplest solution of the Beltrami
equation represents a standing wave with $k^2=\sigma_M^2$ and non-trivial
helicity (\ref{helicity}).

This concludes our brief review of the instability of then chiral medium with $\mu_5\neq 0$
with respect to spontaneous generation of helical classical magnetic fields. As far
as we know, there are no estimates of the life-time of the unstable vacuum, if we start
with vanishing magnetic field.

Another result seems worth mentioning. As can readily be seen, in the limit of vanishing
momentum, $k_i\to 0$, the propagator (\ref{proppole}) tends to: 
\begin{eqnarray}
\label{polenew}
 D_{ij}  ( \vec{k}) \rightarrow -
\frac{i \epsilon_{ijl} k^l}{\sigma_M k^2},~~~\vec{k}\rightarrow0~.
\end{eqnarray}
One can prove \cite{khaidukov} that this result survives with account of all
the electromagnetic corrections.
On the other hand 
the term (\ref{polenew})
results in a topological type interaction of static current loops ($k\ll \sigma_M$), 
proportional to their linking number. Thus, this topological interaction 
of external Wilson loops is not
renormalized.

In this brief review we have illustrated infrared instabilities which are inherent
to macroscopic manifestations of the chiral anomaly. From phenomenological
 point of view of view, the most crucial question is whether the anomaly is relevant
at all. In particular, the electromagnetic fields should be intense and satisfy conditions
like $\sqrt{eB}\gg m_f$ where $m_f$ is the scale of explicit chiral symmetry breaking.
This condition is difficult to satisfy in realistic sets up.

\subsection{Decoherence of fermionic wave function}

From theoretical point of view, it is most intriguing that one predicts \cite{surowka}
the effect of the quantum anomaly to survive in the classical, hydrodynamic limit.
Moreover, the corresponding current (\ref{cme}) is predicted to be dissipation free
since it exists in the equilibrium \cite{cheianov,yee}. 

There is explicit derivation
of the chiral magnetic effect in terms of the Berry phase in the collision-less approximation
to chiral plasma \cite{yamamoto}. However, it seems far from being obvious that
decoherence of the wave function, due to interaction with the medium, does not
damp down the effect of the topological phase. It is worth mentioning in this connection
\footnote{The remark is due to D.E. Kharzeev.}
that there is an analog of the chiral magnetic effect in (1+1) dimensional quantum
wires, see \cite{cheianov} and references therein :
\begin{equation}\label{conductance}
\langle j_{el}\rangle~=~e(\mu_L-\mu_R)~,
\end{equation}
where $\mu_L,\mu_R$ are the chemical potentials for left- and right-movers,
respectively. Eq. (\ref{conductance})
 expresses the so called universal conductance and it works for quantum wires
as far as the wire is not too long, to avoid scattering,
$l_{wire}\ll l_{m.f.p.}$ . One could imagine
that scattering in (3+1) dimensions  plays a similar role. Then one would need, say,
superfluid  to observe the chiral magnetic effect. For discussion of a particular model
of this type see \cite{kirilin}.

 \section{Conclusions}

Thus, we have demonstrated that infrared sensitivities are inherent to chiral media.
Eventually all the infrared effects go back to the fact that chiral media introduce
massless charged particles. The actual mechanism of infrared regularization is not known
and may vary from case to case. In this sense, the physics of the chiral media could be even
richer than one would think.

Mostly, these notes are a kind of a mini-review since many points have already been made
in literature. In conclusion, let us emphasize original points which might be in variance
with conclusions of other papers:
\begin{itemize}
\item{The effective action for photon in chiral media is commonly derived from
the chiral anomaly and in static limit
takes the form of the 3d topological photon mass. In paper \cite{khaidukov} it was demonstrated
however that this mass turns to be imaginary and signals instability of 
magnetostatics of the chiral media.}
\item{Moreover, field theory does not provide any apparent mass scale for the hydrodynamic
expansion for chiral media and, in principle, higher orders could be the same important as
the terms fixed by the anomaly.}
\item{We have demonstrated that standard derivations of the
chiral magnetic effect and chiral separation effect refer in fact to different infrared limits.}
\item{ Chiral effects have been derived in collisionless approximation and reflect topological
contributions to phases of fermionic wave functions. We emphasized that decoherence arising
due to interactions among the constituents could wipe the effects out. One of the ways out
is to concentrate on superfluids.}
\end{itemize}

The authors are grateful to A.S. Gorsky, Z. Khaidukov, D.E. Kharzeev, M. Stephanov for useful
discussions. We are thankful to the organizers of the seminar devoted to the memory of
Academician I.Ya. Pomeranchuk for the invitation to the seminar. The work of VPK and AVS has been supported by grant IK-RU-002 of Helmholtz Association.

\end{document}